\documentclass[a4paper]{JAC2000}
\usepackage{epsf}
\usepackage{daniel}
\title{BEAM LOADING COMPENSATION IN THE MAIN LINAC OF CLIC}
\author{D. Schulte and I. Syratchev, CERN, 1211 Geneva, Switzerland}
\begin{document}
\maketitle
\begin{abstract}
Compensation of multi-bunch beam loading is of great importance in the
main linac of the Compact Linear Collider (CLIC).
The bunch-to-bunch energy variation has to
stay below 1 part in $10^3$.
In CLIC, the RF power is obtained by decelerating a drive beam which is formed
by merging a number of short bunch trains.
A promising scheme 
for tackling beam loading in the main linac is based on
varying the lengths of the bunch trains in the drive beam.
The scheme and its expected performance are presented.
\end{abstract}
\section{Introduction}
Multi-bunch beam loading is a strong effect in the main linac of CLIC.
It needs to be compensated with help of the RF to avoid extreme variations of
the beam energy along the pulse. Several approaches to solve the
problem exist.
All of them are based on manipulations of the drive beam which generates the
RF power.

The first possibility is to reduce the bunch charge in the first part of the
drive-beam pulse~\cite{c:lars}. In this scheme, the first bunch has about
$70\u{\%}$ of the nominal charge. The charge is then slowly increased from one
bunch to the next until it reaches the nominal value. This charge ramp
creates a ramp in the RF voltage. By carefully shaping the charge ramp,
one can achieve beam-loading compensation. In principle, this compensation can
be perfect.
However, it may be very difficult to control the bunch charge with the required
precision so as to achieve the required compensation of the gradient
variation $\Delta G/G_0\le10^{-3}$ ($G_0$ is the nominal
gradient).

Another method is described in reference~\cite{c:beamload}. It achieves
$\Delta G/G_0\approx2\times10^{-3}$. It requires
additional hardware and may compromise the stability of the drive beam
in the decelerator.

In a third option, presented in this paper, one creates a ramp in the
current of the drive-beam pulse
comparable to the first option. But instead of varying the bunch charge, one
varies the number of bunches per unit length of the pulse. This can be
achieved by modifying the drive beam in the drive-beam injector~\cite{c:me}.
To understand this, it is necessary to understand the drive-beam
generation, which is described below.

\section{The Drive-Beam Generation}
\begin{figure}
\centerline{
\epsfxsize=7.8cm
\epsfbox{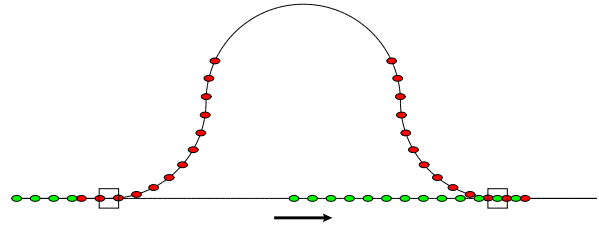}}
\caption{Schematic layout of the delay loop after the drive-beam accelerator
in CLIC. The two RF-deflectors are shown as rectangles.}
\label{c:delay}
\end{figure}
The drive beam is produced and accelerated at a frequency of about
$937\u{MHz}$. In the injector of the drive-beam linac,
one has a sub-harmonic buncher which
can be switched to fill either odd or even buckets.
In the drive-beam accelerator, the beam then consists of short trains of
bunches that fill every second bucket. The first train fills the odd
buckets, the immediately following second train fills the even ones, and this
pattern is then repeated~\cite{c:delay}. The current in the
drive-beam accelerator, and consequently the beam loading, therefore remains
constant.
After acceleration, the trains are separated using an RF-deflector running
at half the linac frequency. The first train is deflected into a delay loop
and merged with the second one in a second RF-deflector, see
Fig.~\ref{c:delay}. The newly created pulses are separated by gaps that allow
conventional deflectors to be switched on and off.

They are sent into two
combiner rings~\cite{c:ring}. These rings have a circumference
equal to the distance between two pulses plus (or minus) a quarter wavelength.
This allows to merge four pulses to form a single one, using an RF-deflector.
The new pulse has four times as many bunches as each of the initial
ones, with a distance between the bunches that is four times smaller.
The bunches comprising the four pulses have been inter-leaved by this
operation so that the
first bunch of each of the initial pulses is one of the first four of the
final pulse.

The first ring is followed by a second one, four times larger, which
merges four of the pulses of the first ring.
At the end, the bunch-to-bunch distance has been reduced from the initial
$64\u{cm}$ to only $2\u{cm}$. In the following, each $64\u{cm}$ long section
of the beam pulse is called a bin and it contains 32 bunches.
The bunches that were in the first bin of
each initial train are in the first bin of the final pulse. The bunches that
were in the second bin of an initial pulse are in the second bin of the final
pulse, and so on.

\section{Delayed Switching}
\begin{figure}
\epsfxsize=7.8cm
\epsfbox{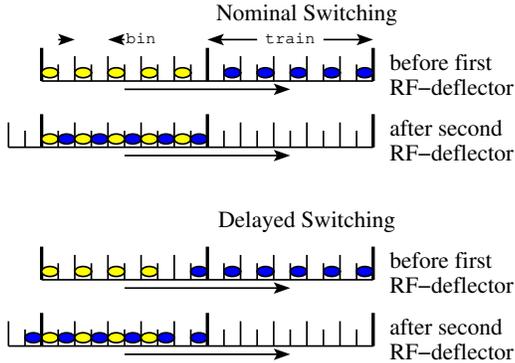}
\caption{The scheme of delayed switching. In this example, each train contains
5 bins (an arbitray number chosen for better visibility).
In the upper case the phase is
switched at the nominal time, creating a rectangular pulse. In the lower case
the phase switch is delayed to create a ramp.
}
\label{f:scheme0}
\end{figure}

In order to create a current ramp in the final pulse,
the first few bins of this pulse must contain a smaller number of bunches
than nominal. This in turn requires that some of the pulses after the
delay loop have less than the nominal two bunches per bin. This can be
achieved by delaying the switching of the sub-harmonic buncher.
The effect of the nominal switching is illustrated in the upper part of
Fig.~\ref{f:scheme0}. The two trains before the delay loop and the
pulse after this loop are shown.
In the delay loop, the bunches of the first train are delayed by one
nominal train length.
In the lower part of the figure,
the sub-harmonic buncher is switched slightly later.
The bunch that, in the nominal scheme,
would have been the first one of the second train is therefore
appended to the first train. The second train starts one bunch later than
nominal.
As a consequence,
the pulse after the delay loop contains only one bunch in the first bin.
The last bunch of the first train is appenended after at the end of the pulse.

The additional tail of the pulse creates no
problem in the combiner ring, as long as the distance to the first bin of the
next pulse is long enough to switch the ejection kickers of the rings on and
off. In the drive-beam decelerator, the additional tail is not
important, since it will just add a little tail to the RF-pulse produced in the
power extraction and transfer structures (PETS).

The switching time can be individually chosen for each train, so a rather fine
ramp in the final pulse can be created.
This solution does not require any additional hardware; one must only be
able to switch the sub-harmonic buncher at non-regular intervals.

\section{Numerical Results}
To achieve beam-loading compensation in the CLIC main linac,
11 of the 32 initial trains need to be delayed in the drive-beam linac.
The maximum delay necessary
is 11 bins. The gradient seen by the main-linac bunches can be simulated
with ASTPC~\cite{c:astpc}.
In this program, the transient effects in the PETS 
of the drive-beam decelerator, as well as in the main-linac
accelerating
structures are taken into account. Each structure is represented by a series of
reflectors that are located at the cell boundaries. This makes it possible to
simulate precisely the beam acceleration in the time domain.
 
\begin{figure}
\centerline{
\epsfxsize=7.8cm
\epsfbox{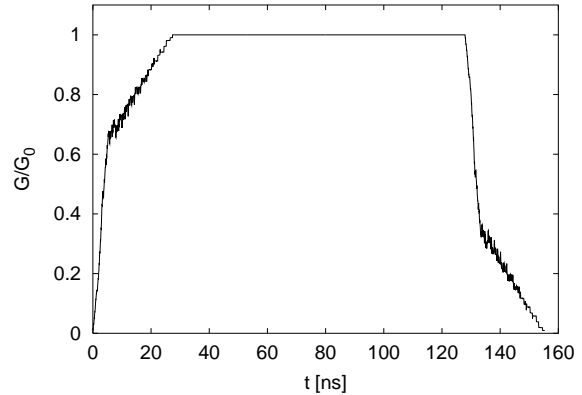}}
\caption{Shape of the RF pulse which is produced by the drive beam if
delayed switching is applied.}
\label{f:res3}
\end{figure}

\begin{figure}
\centerline{
\epsfxsize=7.8cm
\epsfbox{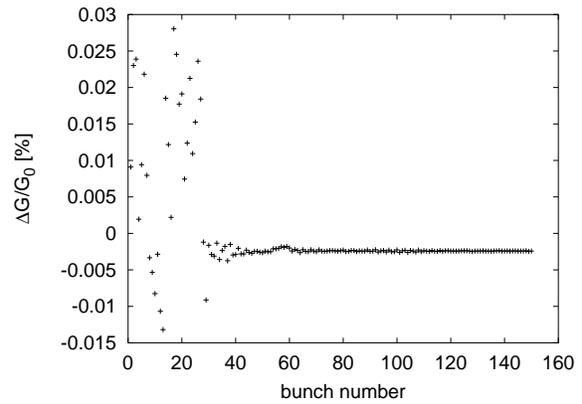}}
\caption{The deviation from the nominal gradient as seen by each bunch in the
main linac.}
\label{f:res1}
\end{figure}

The gradient errors depend on which of the trains are delayed. To find a good
choice, a number of different delay patterns was created randomly.
These were evaluated with the program and the best case was accepted.
For this case,
Fig.~\ref{f:res3} shows the RF-pulse as it is produced by the PETS. This
pulse leads to a bunch-to-bunch gradient error in the main linac that
remains below $\Delta G/G_0=5\times10^{-4}$, see Fig.~\ref{f:res1}. This is
better than the required precision of $\Delta G/G_0\le10^{-3}$.

The method described achieves a constant amplitude of the accelerating field
in the main linac. The main beam is, however, not accelerated on the crest of
the RF wave, but at a small phase in the main part of the linac,
$\Phi_{RF}=6^\circ$. At the
end of the acceleration, this phase is even larger, $\Phi_{RF}=30^\circ$.
Since the amplitude is increased in the RF phase and the beam loading is in
phase with the beam, this leads to an effective phase shift of the total
accelerating field during the first part of the main-linac pulse. In order to
prevent this, one can think of shifting the delayed trains before they
are merged with the other ones. The shift has to be such that the bunches are
in phase with the main beam. In this case, not only the amplitude but also the
acceleration phase is maintained.

\section{Simulation of the Drive Beam}
To estimate the impact of the beam-loading compensation on the stability
of the drive beam in the decelerator, simulations are performed using
PLACET~\cite{c:placet}. A lattice is chosen in which each six-waveguide
structure feeds three main linac structures.

\begin{figure}
\epsfxsize=7.8cm
\epsfbox{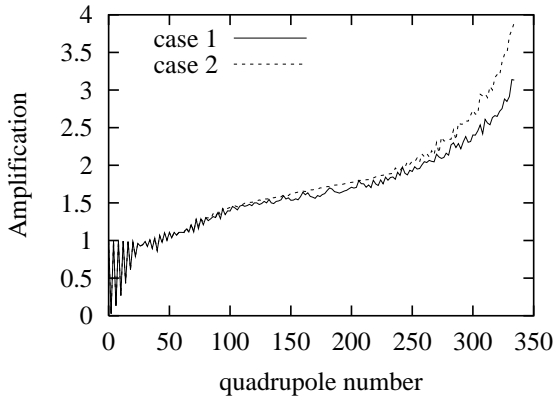}
\caption{The maximum amplification of an initial beam jitter along the
decelerator. The case of a rectangular current pulse (case 1) is compared to
the one with a ramp in bunch number (case 2).}
\label{f:stab}
\end{figure}

As a measure of the stability, the maximum amplification of an initial jitter
is used, which is determined as follows:
in the simulation, each bunch is cut into slices. The beam is offset and then
tracked through the decelerator. The maximum offset that the centre of
any slice reaches, divided by the initial offset, is the maximum amplification.
Figure~\ref{f:stab} shows this amplification of a transverse jitter along the
drive-beam decelerator. If no transverse wakefields were present, the final
amplification factor would be $A=\sqrt{10}$ from the adiabatic undamping of the
motion. As can be seen, a rectangular current pulse (case 1) is close to this
case. The bunch ramp increases the amplification somewhat (case 2).
This seems tolerable. Most of the effect is due to the trailing bunches.

If the delayed trains are shifted in phase, so as to prevent phase shift of the
acceleration field, the wakefield effects in the drive-beam decelerator may
become worse. The simulation shows that also in this case,
the jitter amplification is almost the same as without the shift;
they could not be distinguished in the plot. The method
therefore seems to be practical. But other methods, such as a slow phase change
along the train, might achieve the same result.

\section{Application to CTF3}
\begin{figure}
\centerline{
\epsfxsize=8cm
\epsfbox{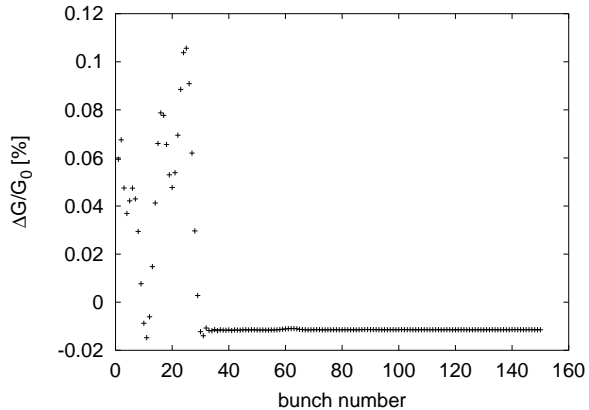}}
\caption{Deviation from the nominal gradient, as seen by the
main-beam bunches in CTF3.}
\label{f:res2}
\end{figure}
Delayed switching could also be used in CTF3, the new CLIC Test Facility,
which will be constructed at CERN.
In this case, the switching time
will be longer, about $4\u{ns}$. Since only ten pulses are merged to form the
drive beam, one only delays three of them.
Again, different cases were searched for
an optimum. The achieved compensation is very good,
about $\Delta G/G_0\approx1.2\times10^{-3}$, see Fig.~\ref{f:res2}.

\section{Conclusion}
The method presented, to compensate the beam loading in the main linac,
achieves the required precision of better than one part in 1000. It is very
simple, can be adjusted to different switching times, and requires no
additional hardware. It seems to be the method of choice for CLIC.

\end{document}